\documentclass[prd,aps,draft]{revtex4}

\begin{document}
%%%%%%%%%%%%%%%%%%%%%%%%%%%%%%%

\title{Detecting ill posed boundary conditions in General Relativity}

\author{Gioel Calabrese and Olivier Sarbach}

\affiliation{Department of Physics and Astronomy, Louisiana State
University, 202 Nicholson Hall, Baton Rouge, Louisiana 70803-4001}

%%%%%%%%%%%%%%%%%%%%%%%%%%%%%%%%
\begin{abstract}
%%%%%%%%%%%%%%%%%%%%%%%%%%%%%%%%

A persistent challenge in numerical relativity is the correct
specification of boundary conditions. In this work we consider a many
parameter family of symmetric hyperbolic initial-boundary value
formulations for the linearized Einstein equations and analyze its
well posedness using the Laplace-Fourier technique. By using this
technique ill posed modes can be detected and thus a necessary
condition for well posedness is provided. We focus on the following
types of boundary conditions: i) Boundary conditions that have been
shown to preserve the constraints, ii) boundary conditions that result
from setting the ingoing constraint characteristic fields to zero and
iii) boundary conditions that result from considering the projection
of Einstein's equations along the normal to the boundary
surface. While we show that in case i) there are no ill posed modes,
our analysis reveals that, unless the parameters in the formulation
are chosen with care, there exist ill posed constraint violating modes
in the remaining cases.
\end{abstract}

\maketitle
%%%%%%%%%%%%%%%%%%%%%%%%%%%%%%%%%%%%%%%%%%%%%%%%%%%%%%%%%%%%%%
\section{Introduction}
%%%%%%%%%%%%%%%%%%%%%%%%%%%%%%%%%%%%%%%%%%%%%%%%%%%%%%%%%%%%%%

Obtaining a long time convergent numerical simulation of a binary
black hole spacetime in domains with artificial boundaries continues
to be a challenge in numerical relativity and one which has recently
received a substantial amount of attention, notably in the case of
hyperbolic formulations (see \cite{R,L} for reviews). The challenge
remains in part because of the difficulty in specifying boundary
conditions.  It has been recognized
\cite{FN,Stewart,IR,CLT,CPSTR,SSW,SW,FG1,FG2,BB} that the boundary
conditions have to satisfy two important requirements. First, they
have to preserve the constraints. By this we mean that they must
guarantee that if the constraints are satisfied initially they are
also satisfied at later times. We refer to boundary conditions that
satisfy this property as constraint preserving boundary conditions
(CPBC). Second, the boundary conditions have to be such that the
resulting initial-boundary value problem (IBVP) is well posed. This
means that given initial and boundary data a unique solution exists
and that at each fixed time the solution depends continuously on the
data. Well posedness is a necessary condition for the construction of
consistent and stable finite difference schemes
\cite{GKO-Book,CPST-Convergence}.

When the evolution equations are in symmetric hyperbolic form one
usually specifies maximal dissipative boundary conditions
\cite{LaxPh}. Under certain technical assumptions, these conditions
guarantee that the resulting IBVP is well posed \cite{Secchi,
Rauch}. Using maximal dissipative boundary conditions, Friedrich and
Nagy \cite{FN} were able to find well posed CPBC for a particular
formulation of the full nonlinear vacuum equations. However, most of
the hyperbolic formulations used in numerical relativity are based on
evolution equations that use a different set of variables than in
Ref. \cite{FN}. For these formulations, the derivation of well posed
CPBC seems to be more difficult. Part of the problem stems from the
fact that CPBC result in a set of partial differential equations that
must hold at the boundary surface, and it is not always possible to
cast these equations into the form of maximal dissipative boundary
conditions.  This is probably the reason why current well posed CPBC
for formulations other than that used in Ref. \cite{FN} are either
limited to homogeneous boundary data \cite{SW} or to linearizations
around a Minkowski background \cite{CPSTR,SSW,SW}. Even in those
cases, the CPBC obtained so far might be too restrictive in the sense
that they do not allow the specification of the physical quantities at
the boundary with the freedom one would like to have. For example, the
well posed boundary conditions obtained in Ref. \cite{CPSTR} involve a
coupling between the in- and outgoing variables and, likely, this
coupling will introduce reflections at the boundary. Therefore, more
general techniques are desirable in order to show well posedness for
more generic CPBC.

In this article, we use the Laplace-Fourier technique to analyze
boundary conditions in linearized General Relativity. This technique
is very useful when the evolution equations are linear and have
constant coefficients since it can be applied to boundary conditions
that are more general than the maximal dissipative ones.
Specifically, it can be applied to boundary conditions which have the
form of differential equations at the boundary.  Furthermore, the
method is capable of detecting the presence of ill posed modes
analytically. Ill posed modes are solutions to the IBVP that grow
exponentially in time with an exponential factor that can be
arbitrarily large, and their existence makes it impossible for the
solution to depend continuously on the data. The Laplace-Fourier
technique therefore provides us with a necessary condition for well
posedness. However, it should be emphasized that the absence of ill
posed modes (as defined in this article) does not automatically
guarantee well posedness.  Although more complicated in this case,
results for the variable coefficient case are available by freezing
the coefficients at the boundary (see \cite{Kreiss,MO}).

This article is organized as follows: The conditions under which the
specification of non-maximal dissipative boundary conditions for
symmetric hyperbolic systems with constant coefficients yields ill
posed modes are reviewed in section \ref{eigenvalues}. In section
\ref{CPBC} we discuss the boundary conditions that have been
considered for the generalized Einstein-Christoffel formulation of
Einstein's equations \cite{KST, AY} when linearized around flat
spacetime. The generalized Einstein-Christoffel system is a family of
symmetric hyperbolic formulations that is parametrized by a constant
$\eta$. The boundary conditions we are considering are: i) The CPBC
that were considered in Ref. \cite{CPSTR} and that are based on
solving a closed evolution system at the boundary and on maximal
dissipative boundary conditions. ii) Boundary conditions that are
obtained by setting the ingoing constraint characteristic fields to
zero. iii) Boundary conditions that are obtained by considering the
projection of Einstein's equations along the normal to the boundary
surface, as recently proposed by Frittelli and Gomez
\cite{FG1,FG2}. In section \ref{LFA} we apply the techniques discussed
in section \ref{eigenvalues} and show that the cases ii) and iii)
suffer from the presence of ill posed modes unless the parameter
$\eta$ in the generalized Einstein-Christoffel formulation lies in a
specific range. We also show that there are no ill posed modes in case
i) which is consistent with the well posedness estimates derived in
Ref. \cite{CPSTR}. In section \ref{CV} we show that the ill posed
modes we have found in cases ii) and iii) do, in fact, all violate the
constraints. This means that the evolution system for the constraint
variables is ill posed in those cases. Since this system is always
strongly hyperbolic and since our boundary conditions are constructed
from specifying maximal dissipative conditions for this system, this
also illustrates that maximal dissipative boundary conditions do not
necessarily yield a well posed formulation if the evolution equations
are strongly hyperbolic (but not symmetrizable). In particular, our
calculations show that the boundary conditions that are constructed
following the schemes ii) and iii) do not necessarily lead to CPBC and
that one should always check the evolution system for the constraint
variables. Our results and their implications on deriving well posed
CPBC are discussed in section \ref{conclusions}. A similar analysis
for the Frittelli-Reula \cite{FR} system has been undertaken by
Stewart \cite{Stewart}.

%%%%%%%%%%%%%%%%%%%%%%%%%%%%%%%%%%%%%%%%%%%%%%%%%%%%%%%
\section{Detecting ill posed modes}
\label{eigenvalues}
%%%%%%%%%%%%%%%%%%%%%%%%%%%%%%%%%%%%%%%%%%%%%%%%%%%%%%%%

In this section, we review the techniques that can reveal the presence
of ill posed modes. They are based on a Laplace transformation in time
and on a Fourier transformation in the spatial directions that are
tangential to the boundaries and are described in more detail in
Refs. \cite{KL-Book,GKO-Book}. For simplicity, we restrict the
following discussion to the 2D case; the generalization to 3D is
straightforward.

Consider a 2D first order in time and space linear evolution
equation of the form
\begin{equation}
\partial_t u = A\partial_x u + B\partial_y u,
\label{Eq:MainEqs}
\end{equation}
where $u = u(t,x,y)$ is a vector-valued function and the
matrices $A$ and $B$ are constant and symmetric. We consider
solutions to Eq. (\ref{Eq:MainEqs}) on the domain $t > 0$, $x > 0$,
$-\pi < y < +\pi$ with initial data
\begin{equation}
u(0,x,y) = f(x,y)
\label{Eq:IniData}
\end{equation}
and boundary conditions at $x=0$ of the form
\begin{equation}
L(\partial_t,\partial_y)u(t,0,y) = g(t,y),
\label{Eq:BData}
\end{equation}
where $L$ is a linear operator with constant coefficients that only
involves derivatives which are tangential to the boundary. For
technical reasons, we assume that $L(\partial_t,\partial_y)$ is
homogeneous in the sense that $L(\mu\partial_t,\mu\partial_y)
=\mu L(\partial_t,\partial_y)$ for all positive $\mu$. We also
assume periodic boundary conditions in the $y$-direction (similar
conclusions hold for the case $-\infty < y < +\infty$).

The IBVP (\ref{Eq:MainEqs}), (\ref{Eq:IniData}), (\ref{Eq:BData}) is
said to be well posed\footnote{When discussing initial-boundary value
problems one sometimes demands a slightly stronger estimate that also
bounds the norm of the solution at the boundary surface. For the
purpose of the present article, it is sufficient to consider the
weaker estimate (\ref{Eq:WP}) since we will show that when the
determinant condition is violated, the inequality (\ref{Eq:WP}) cannot
hold for all initial data}, if given smooth square integrable data
$f$, $g$ there exists a unique smooth solution. Furthermore, there are
constants $C$, $a$ such that
\begin{equation}
\| u(t,.) \|^2 \leq C e^{at} \left[ \| f \|^2 + \int_0^t \| g(\tau,.) \|^2 d\tau \right],
\label{Eq:WP}
\end{equation}
for all $t > 0$ and all square integrable data $f$ and $g$. Here, $\|
u(t,.)  \|$ denotes the $L^2$ norm of $u$ defined by $\| u(t,.)\|^2 =
\int_{x>0} |u(t,x,y)|^2 dxdy$ and similarly, $\| f \|^2 = \int_{x>0}
|f(x,y)|^2 dxdy$ and $\| g(\tau,.) \|^2 = \int | g(\tau,y) |^2
dy$. The estimate (\ref{Eq:WP}) implies that for each fixed $t$, the
solution depends continuously on the data $f$ and $g$.

A first step in checking if a given initial-boundary formulation
satisfies a well posedness inequality of the type (\ref{Eq:WP}) is to
look for solutions of the problem with homogeneous data ($g=0$) which
are of the form
\begin{equation}
u(t,x,y) = e^{st + i\omega y} \tilde{u}(x),
\label{Eq:SS}
\end{equation}
where $\omega$ is an integer, $s$ is complex with $\mbox{Re}(s) > 0$
and $\tilde{u}(x)$ is a smooth function that lies in
$L^2(0,\infty)$. If such a solution exists, the problem cannot be well
posed. In order to see this we notice that the functions
\begin{equation}
u_{m}(t,x,y) = e^{m(st + i\omega y)} \tilde{u}(m x),
\end{equation}
where $m = 1,2,3,...$ can be arbitrarily large are also solutions and
since $\| u_{m}(t,.) \|/\| u_{m}(0,.)\| = \exp(m\mbox{Re}(s) t)$ the
estimate (\ref{Eq:WP}) cannot hold with constants $C$ and $a$ that are
independent of the initial data. Therefore, an obvious check for well
posedness is to see whether or not
Eqs. (\ref{Eq:MainEqs}), (\ref{Eq:BData}) admit nontrivial solutions of
the form (\ref{Eq:SS}) with homogeneous boundary data.

Using expression (\ref{Eq:SS}) in Eqs. (\ref{Eq:MainEqs}),
(\ref{Eq:BData}), we obtain (for $g = 0$)
\begin{eqnarray}
&& s\tilde{u} = A\partial_x\tilde{u} + i\omega B\tilde{u},
\label{Eq:EV1}\\
&& L(s,i\omega)\tilde{u}(0) = 0.
\label{Eq:EV2}
\end{eqnarray}
These equations form a system of ordinary differential equations and
can be solved analytically. In order to do so, we first bring $A$ to
diagonal form through an orthonormal transformation. Thus, the matrix
$B$ is still symmetric and we can write
\begin{displaymath}
A = \left( \begin{array}{cc}  0 & 0 \\ 0 & A_1 \end{array} \right),
\end{displaymath}
where $A_1 = \mbox{diag}(\lambda_1,...,\lambda_p,\lambda_{p+1},...,\lambda_{p+q})$
with $\lambda_1,...\lambda_p$ real and negative and
$\lambda_{p+1},...,\lambda_{p+q}$ real and positive. Here, $p$ and $q$ 
are the number of in- and outgoing modes, respectively.
Accordingly, we write
\begin{displaymath}
B = \left( \begin{array}{cc}  B_{00} & B_{01} \\ B_{10} & B_{11} \end{array} \right),
\qquad
L = ( L_0, L_1 ), \qquad
\qquad
\tilde{u} = \left( \begin{array}{c} \tilde{u}_0 \\ \tilde{u}_1 \end{array} \right).
\end{displaymath}
Now the zero components of Eq. (\ref{Eq:EV1}) yield the following algebraic
relation between $\tilde{u}_0$ and $\tilde{u}_1$:
\begin{equation}
S_{00}\tilde{u}_0 = -S_{01}\tilde{u}_1\, ,
\end{equation}
where we have introduced the matrix $S = S(s,\omega) = s I - i\omega B$.
Since the matrix $B$ is symmetric, the matrix $S_{00} = s I - i\omega B_{00}$ 
is invertible for all $\mbox{Re}(s) > 0$ and all integer $\omega$,
and we can express $\tilde{u}_0$ in terms of $\tilde{u}_1$:
\begin{equation}
\tilde{u}_0 = -S_{00}^{-1} S_{01}\tilde{u}_1\, .
\end{equation}
Inserting this into the remaining equations of the system
(\ref{Eq:EV1}), (\ref{Eq:EV2}), we obtain the reduced problem
\begin{eqnarray}
\partial_x\tilde{u}_1 &=& M(s,\omega)\tilde{u}_1\, ,
\label{Eq:reduced1} \\
\tilde{L}\tilde{u}_1 &=& 0,
\label{Eq:reduced2}
\end{eqnarray}
where
\begin{eqnarray}
M(s,\omega) &=& A_1^{-1}\left( S_{11}(s,\omega) - S_{10} S_{00}^{-1} S_{01}(s,\omega) \right), 
\nonumber\\
\tilde{L}(s,\omega) &=& L_1 - L_0 S_{00}^{-1} S_{01}.
\nonumber
\end{eqnarray}
One can show \cite{MO} that for $\mbox{Re}(s) > 0$ the matrix
$M(s,\omega)$ has exactly $p$ eigenvalues with negative real parts and
exactly $q$ eigenvalues with positive real parts (the eigenvalues are
counted according to their algebraic multiplicity.)

The eigenvalues of $M$ that have positive real part lead to
exponential growth in $x$. Since the solution $\tilde{u}$ has to be in
$L^2(0,\infty)$, the integration constants have to be chosen such that
there is no such growth in $x$. In order to achieve this, we choose,
for each $(s,\omega)$, a unitary matrix $U = U(s,\omega)$ that brings
$M(s,\omega)$ in upper triangular form:
\begin{equation}
U(s,\omega)^{-1} M(s,\omega) U(s,\omega) = 
\left( \begin{array}{cc} M_-(s,\omega) & M_0(s,\omega) \\ 0 & M_+(s,\omega) \end{array} \right).
\label{Eq:TransM} 
\end{equation}
Here, $M_-$ ($M_+$) is an upper triangular matrix whose eigenvalues
have negative (positive) real parts. If we introduce the new variable 
$v(x) = U(s,\omega)^{-1} \tilde{u}_1(x)$, system (\ref{Eq:reduced1}),
(\ref{Eq:reduced2}) becomes
\begin{eqnarray}
&& \partial_x v_-(x) = M_-(s,\omega) v_-(x) + M_0(s,\omega) v_+(x)\; ,
\nonumber\\
&& \partial_x v_+(x) = M_+(s,\omega) v_+(x)\; ,
\nonumber\\
&& L_- v_-(0) + L_+ v_+(0) = 0,
\nonumber
\end{eqnarray}
where $(L_-,L_+) = \tilde{L} U$. It follows that $v_+$ must vanish for
$v$ to be in $L^2(0,\infty)$. This implies that $v_-(x) = \exp(M_-
x)\sigma_-$ where $\sigma_-$ has to satisfy the boundary condition
$L_-\sigma_- = 0$. We conclude that the system
(\ref{Eq:EV1},\ref{Eq:EV2}) has only the trivial solution if and only
if the determinant condition\footnote{This condition is weaker than
the uniform Kreiss condition \cite{Kreiss} that requires that $| \det
L_- |$ must be bounded away from zero.  The reason why here we do not
require the uniform Kreiss condition is that it might be too strong
for the case of CPBC in General Relativity.  As we will see in section
\ref{LFA} the well posed CPBC that were derived in Ref. \cite{CPSTR}
do not satisfy the uniform Kreiss condition.}
\begin{equation}
\det L_-(s,\omega) \neq 0, \qquad \mbox{Re}(s) > 0,
\label{Eq:Det}
\end{equation}
is satisfied. (In particular, $L_-$ must be a square matrix of
dimension $p$. This means that we need exactly as many independent
boundary conditions as there are ingoing modes). If the determinant
condition is violated at some point $(s,\omega) = (s_0,\omega_0)$, it is
also violated for $(s,\omega) = m(s_0,\omega_0)$ with $m=1,2,3,...$
and the initial-boundary formulation admits solutions of the form
(\ref{Eq:SS}) that grow exponentially in time where the exponential
factor $s$ can have arbitrarily large real part.

In section \ref{LFA} we will discuss the determinant condition for the case
of the linearized Einstein equations with boundaries.

%%%%%%%%%%%%%%%%%%%%%%%%%%%%%%%%%%%%%%%%%%%%%%%%%%%%%%%%%%%%%%
\section{Boundary conditions for the linearized Einstein-Christoffel system}
\label{CPBC}
%%%%%%%%%%%%%%%%%%%%%%%%%%%%%%%%%%%%%%%%%%%%%%%%%%%%%%%%%%%%%%

In this section we discuss boundary conditions for a linearization of the 
generalized Einstein-Christoffel vacuum equations \cite{KST}.  This
formulation has the attractive feature that when linearized around
flat spacetime written in Minkowski coordinates it simply reduces to a
set of six wave equations, written in first order form:
\begin{eqnarray}
\partial_t K_{ij} &=& -\delta^{kl}\partial_k f_{lij} \, , 
\label{Eq:Kij}\\
\partial_t f_{kij} &=& -\partial_k K_{ij} \, .
\label{Eq:fkij}
\end{eqnarray}
Here, $K_{ij}$ denotes the linearized extrinsic curvature and the
symbols $f_{kij}$ represent linear combinations of the linearized
Christoffel symbols $\Gamma_{kij}$:
\begin{equation}
f_{kij} = \Gamma_{(ij)k} + \delta^{rs}\left( \delta_{ki}\Gamma_{[sj]r} + \delta_{kj}\Gamma_{[si]r}
 + \frac{\eta-4}{2\eta}\, \delta_{ij} \Gamma_{[sk]r} \right).
\end{equation}
The value of $\eta$ (which has to be different from zero)
parametrizes the family of formulations. The particular case with
$\eta = 4$ corresponds to the original Einstein-Christoffel system
derived by Anderson and York \cite{AY}. We set the shift to zero, and
the lapse is linearized in such a way that it satisfies the densitized
lapse gauge condition $\alpha = \sqrt{g}$ up to second order
corrections, where $g$ denotes the determinant of the three metric. A
solution to the system (\ref{Eq:Kij}), (\ref{Eq:fkij}) is a solution
to the linearized Einstein equations if and only if the constraints
are satisfied. In terms of the constraint variables
\begin{eqnarray}
C &=& \frac{\eta}{4}\, \delta^{rs}\partial_r v_s\; ,
\label{Eq:Ham}\\
C_j &=& \delta^{rs}\left( \partial_r K_{sj} - \partial_j K_{rs} \right),
\label{Eq:Mom}\\
C_{lkij} &=& 2\partial_{[l} f_{k]ij} + \eta\,\partial_{[l} \delta_{k](i} v_{j)} 
 + \frac{\eta-4}{4}\,\delta_{ij}\partial_{[l} v_{k]}\, , 
\label{Eq:Four}
\end{eqnarray}
where $v_k = \delta^{ij}( f_{kij} - f_{ijk})$, the constraints
are given by $C=0$, $C_j=0$, $C_{lkij} = 0$.

We consider the evolution system (\ref{Eq:Kij}), (\ref{Eq:fkij}) on the
domain $t > 0$, $x > 0$, $-\pi < y,z, < +\pi$ and introduce the
characteristic variables in the $x$ direction\footnote{Notice that the
characteristic variables defined here are related with the ones
$v_{ij}^{(\pm)}$ defined in Ref. \cite{CPSTR} according to
$u_{ij}^{(\pm)} = v_{ij}^{(\mp)}/\sqrt{2}$.}
\begin{eqnarray}
u_{ij}^{(-)} &=& \frac{1}{\sqrt{2}}\left( K_{ij} + f_{xij} \right),\\
u_{ij}^{(+)} &=& \frac{1}{\sqrt{2}}\left( K_{ij} - f_{xij} \right),\\
u_{Aij}^{(0)} &=& f_{Aij}\; .
\label{Eq:DefCVMain}
\end{eqnarray}
Here and in the following capital Latin indices stand for the
tangential directions $y$ and $z$. When written in terms of these
variables the evolution equations (\ref{Eq:Kij}), (\ref{Eq:fkij}) take
the form
\begin{eqnarray}
\partial_t u_{ij}^{(-)} &=& - \partial_x u_{ij}^{(-)} - 
\frac{1}{\sqrt{2}}\delta^{AB}\partial_A u_{Bij}^{(0)}\; ,
\label{Eq:v-ij}\\
\partial_t u_{ij}^{(+)} &=& + \partial_x u_{ij}^{(+)} - 
\frac{1}{\sqrt{2}}\delta^{AB}\partial_A u_{Bij}^{(0)}\; ,
\label{Eq:v+ij}\\
\partial_t u_{Aij}^{(0)} &=& -\frac{1}{\sqrt{2}}\partial_A 
\left( u_{ij}^{(-)} + u_{ij}^{(+)} \right),
\label{Eq:vkij}
\end{eqnarray}
and we see that the matrix $A$ in Eq. (\ref{Eq:MainEqs}) is diagonal.

For the constraints to be satisfied everywhere, when boundaries are
present, one has to ensure that they are satisfied initially and that
no constraint violating mode enters the domain. In order to ensure
this, we follow the analysis in Ref. \cite{CPSTR} and first consider
the evolution of the constraint variables with respect to the flux
defined by the main evolution equations
(\ref{Eq:Kij}), (\ref{Eq:fkij}). One can show \cite{CPSTR} that the
traceless part of $C_{lkij}$ is constant in time, while the remaining
constraints propagate according to
\begin{eqnarray}
\partial_t C &=& \frac{\eta}{4}\, \delta^{ij}\partial_i C_j\; ,
\label{Eq:C}\\
\partial_t C_j &=& \frac{4 - 2\eta}{\eta}\, \partial_j C - \delta^{rs}\partial_r T_{sj}\; ,
\label{Eq:Cj}\\
\partial_t T_{ij} &=& -\partial_i C_j + \left( 1 - \frac{3\eta}{4} \right) \partial_j C_i 
 + \frac{\eta}{4}\, \delta_{ij}\delta^{rs}\partial_r C_s\; ,
\label{Eq:Tij}\\
\partial_t V_{ij} &=& \left( \frac{7\eta}{4} - 3 \right) \partial_{[i} C_{j]}\; ,
\label{Eq:Vij}
\end{eqnarray}
where $T_{ij} = \delta^{rs}( C_{rijs} + C_{ijrs} )$, and $V_{ij} = \delta^{rs}C_{ijrs}$.
Introducing $\kappa = 1 - 3\eta/4$ and the variables
\begin{equation}
C_{ij} = T_{ij} + \frac{2\eta - 4}{\eta}\,\delta_{ij} C
 = \delta^{rs}( \partial_r f_{ijs} - \partial_j f_{irs}) 
 + \kappa\partial_j v_i - \kappa\delta_{ij}\delta^{rs}\partial_r v_s\, ,
\label{Eq:Two}
\end{equation}
the characteristic fields can be written as\footnote{For $\eta=2$ or
$\eta=8/3$ these fields are not complete. When $\eta=2$ it follows
that $C_{ij}$ is traceless and as a consequence,
$\delta^{AB}V_{AB}^{(0)} = 0$. But in this case one has the additional
field $2C + C_{xx}$ that propagates with zero speed. When $\eta=8/3$,
$\tilde{V}_{ij} = 0$ can be replaced by the fields $V_{AB}$ and $6V_{xA} +
5C_{xA}$.}
\begin{eqnarray}
V_j^{(-)} &=& \frac{1}{\sqrt{2}}\left( C_j + C_{xj} \right), \\
V_j^{(+)} &=& \frac{1}{\sqrt{2}}\left( C_j - C_{xj} \right), \\
V_{Aj}^{(0)} &=& C_{Aj} + \kappa\left( \delta_{xj} C_{xA} - \delta_{Aj} C_{xx} \right),\\
\tilde{V}_{ij}^{(0)} &=& -\frac{7\kappa + 2}{3}C_{[ij]} + (\kappa+1)V_{ij}\, .
\end{eqnarray}
If the system (\ref{Eq:C}), (\ref{Eq:Cj}), (\ref{Eq:Tij}), (\ref{Eq:Vij}) is symmetric (or
symmetrizable) hyperbolic, one can guarantee that if the constraints
are satisfied initially and if homogeneous maximal dissipative
boundary conditions are given, the constraints will be satisfied
everywhere. Therefore, we consider boundary conditions at
$x=0$ which are of the form
\begin{eqnarray}
&&V_x^{(-)} - a V_x^{(+)} = 0, \qquad
V_A^{(-)} - b V_A^{(+)} = 0,
\label{Eq:CPBC}\\
&&u_{xx}^{(-)} - c\, u_{xx}^{(+)} = g_{xx}, \qquad
\hat{u}_{AB}^{(-)} - d\, \hat{u}_{AB}^{(+)} = \hat{g}_{AB}, \qquad
\label{Eq:Free}
\end{eqnarray}
where the magnitudes of $a$, $b$, $c$ and $d$ are smaller or equal to
$1$ and $\hat{u}_{AB}^{(\mp)} = u_{AB}^{(\mp)} -
\frac{1}{2}\delta_{AB}\delta^{CD} u_{CD}^{(\mp)}$ denotes the
traceless part of $u_{AB}^{(\mp)}$. In order to express the conditions
(\ref{Eq:CPBC}) in terms of the main variables $K_{ij}$, $f_{kij}$, we
use the definition of the constraint variables,
Eqs. (\ref{Eq:Ham}), (\ref{Eq:Mom}), (\ref{Eq:Two}), and the evolution
equations (\ref{Eq:Kij}), (\ref{Eq:fkij}) in order to trade
$x$-derivatives by time and tangential derivatives:
\begin{eqnarray}
V_x^{(\mp)} &=& \delta^{AB}\left[ \pm\partial_t u_{AB}^{(\mp)} + 
\partial_A u_{xB}^{(\mp)} \pm\frac{\delta^{CD}}{\sqrt{2}}\, \partial_C u_{DAB}^{(0)}
 \pm\frac{\kappa}{2}\partial_A
\left( \sqrt{2}\delta^{CD} u_{CDB}^{(0)} - \sqrt{2}\delta^{ij} u_{Bij}^{(0)} + u_{xB}^{(-)} - u_{xB}^{(+)} \right) \right], 
\nonumber\\
V_A^{(\mp)} &=& \mp\partial_t u_{xA}^{(\mp)} 
 + \delta^{CD}\partial_C u_{AD}^{(\mp)} - \delta^{ij}\partial_A u_{ij}^{(\mp)}
 \mp \frac{\delta^{CD}}{\sqrt{2}}\partial_C u_{DxA}^{(0)}
 \mp\frac{\kappa}{2}\delta^{CD}\partial_A\left( \sqrt{2}u_{CDx}^{(0)} - u_{CD}^{(-)} + u_{CD}^{(+)} \right)\; .
\nonumber
\end{eqnarray}
It follows from the energy estimates derived in Ref. \cite{CPSTR} that
when $0 < \eta < 2$ the conditions (\ref{Eq:CPBC}) guarantee that the
constraints are satisfied everywhere if they are satisfied
initially. In the following, we will also consider other values of
$\eta$ and show that one might have ill posed modes if the parameter
$\eta$ lies outside the interval $(0,2)$.  Notice that the conditions
(\ref{Eq:CPBC}) do not involve derivatives normal to the boundary
($\partial_x$). They can be interpreted as evolution equations for
the variables $\delta^{AB}(u_{AB}^{(-)} + a u_{AB}^{(+)})$ and
$u_{xA}^{(-)} + b u_{xA}^{(+)}$ at the boundary.  The functions
$g_{xx}$ and $\hat{g}_{AB}$ are data that can be given freely for a
combination of the in- and outgoing gauge and physical variables,
respectively \cite{CPSTR}.

In the next section, we will analyze the following choices of
parameters:
\begin{enumerate}
\item $a=-1$, $b=1$, $c=d=1$:\\
 This corresponds to the Neumann boundary conditions that we have
discussed in Ref. \cite{CPSTR}. In this case the boundary conditions
can be recast in a closed evolution system at the boundary. Its
solutions provide boundary data for the main evolution system in the
form of maximal dissipative boundary conditions. When $0 < \eta < 2$
one can derive well posedness estimates for the resulting IBVP and the
boundary conditions can indeed be called CPBC.
\item $a=1$, $b=-1$, $c=d=-1$:\\
This corresponds to the Dirichlet conditions specified in Ref. 
\cite{CPSTR}.  They can also be recast in a closed evolution system at
the boundary, and for $0 < \eta < 2$ one has a well-posed IBVP
with CPBC.
\item $a=0$, $b=0$:\\
This corresponds to setting the ingoing constraint variables to zero
and might be the most obvious choice for obtaining CPBC. However, we
will show in the next section that the resulting IBVP possesses
ill posed modes unless the parameter $\eta$ is chosen appropriately.
\item $a=0$, $b=1$:\\ These are the conditions that one obtains after
linearizing the boundary conditions that were recently proposed in
Ref. \cite{FG2}. There, the Einstein-Christoffel formulation ($\eta =
4$) is considered and the boundary conditions (\ref{Eq:CPBC}) are
obtained by projecting Einstein's equations along the normal to the
boundary rather than by analyzing the evolution of the constraints. In
fact, one can show that setting $G_{xA}$ to zero and
rewriting\footnote{Actually this procedure is not unique since there
is an ambiguity when first derivatives of the variables $K_{ij}$ and
$f_{kij}$ are substituted for second derivatives of the
three-metric. This ambiguity stems from the fact that one can always
change the resulting expression by using the constraints
$C_{lkij}=0$. For definiteness, we take the choice that leads to the
same boundary conditions as in Ref. \cite{CPSTR}.} the resulting
equations in terms of the variables $K_{ij}$ and $f_{kij}$ is
equivalent to the second equation in (\ref{Eq:CPBC}) with $b=1$, while
setting $G_{xx}$ ($G_{tx}$) to zero is equivalent to the first
equation in (\ref{Eq:CPBC}) with $a=1$ ($a=-1$). In Ref. \cite{FG2},
the authors propose to set the linear combination $G_{xx}-G_{tx}$ to
zero which would correspond to using $a=0$ in (\ref{Eq:CPBC}). In the
next section, we show that the resulting boundary conditions yield an
ill-posed formulation if the parameter $\eta$ is not chosen
appropriately.
\end{enumerate}

%%%%%%%%%%%%%%%%%%%%%%%%%%%%%%%%%%%%%%%%%%%%%%%%%%%%%%%%%%%%%%
\section{Laplace-Fourier analysis}
\label{LFA}
%%%%%%%%%%%%%%%%%%%%%%%%%%%%%%%%%%%%%%%%%%%%%%%%%%%%%%%%%%%%%%

Following the analysis described in section \ref{eigenvalues}, we look
for solutions to Eqs. (\ref{Eq:Kij}), (\ref{Eq:fkij}),
(\ref{Eq:CPBC}), (\ref{Eq:Free}) with homogeneous boundary data
($g_{xx} = 0$, $\hat{g}_{AB} = 0$) and which are of the form
\begin{eqnarray}
u_{ij}^{(\mp)}(t,x,y,z) &=& e^{st + i\omega_y y + i\omega_z z} \tilde{u}_{ij}^{(\mp)}(x),\label{Eq:SSEC1}\\
u_{Aij}^{(0)}(t,x,y,z) &=& e^{st + i\omega_y y + i\omega_z z} \tilde{u}_{Aij}^{(0)}(x),\label{Eq:SSEC2}
\end{eqnarray}
where $s$ is a complex number with positive real part and $\omega_y$
and $\omega_z$ are integers. For the solution to be square integrable, we
require the functions $\tilde{u}_{ij}^{(\mp)}(x)$ and
$\tilde{u}_{Aij}^{(0)}(x)$ to be in $L^2(0,\infty)$.  From Eqs.
(\ref{Eq:SSEC1}), (\ref{Eq:SSEC2}) and (\ref{Eq:vkij}) we obtain an
algebraic condition 
\begin{equation}
s\tilde{u}_{Aij}^{(0)} = -\frac{i}{\sqrt{2}} \omega_A 
 \left( \tilde{u}_{ij}^{(-)} + \tilde{u}_{ij}^{(+)} \right)
\label{Eq:zeromode}
\end{equation}
that can be used to eliminate the variable $\tilde{u}_{Aij}^{(0)}$
from the remaining equations.  Inserting Eqs. (\ref{Eq:SSEC1}),
(\ref{Eq:SSEC2}), (\ref{Eq:zeromode}) into
Eqs. (\ref{Eq:v-ij}), (\ref{Eq:v+ij}) yields the ordinary differential
equation
\begin{equation}
\partial_x \left( \begin{array}{c} \tilde{u}_{ij}^{(-)} \\ \tilde{u}_{ij}^{(+)} \end{array} \right)
 = M(s,\underline{\omega}) \left( \begin{array}{c} \tilde{u}_{ij}^{(-)} \\ \tilde{u}_{ij}^{(+)} \end{array} \right),
\label{Eq:ODE}
\end{equation}
where
\begin{equation}
M(s,\underline{\omega}) = 
\left( \begin{array}{cc} -s - \frac{\underline{\omega}^2}{2s} & -\frac{\underline{\omega}^2}{2s} \\ 
\frac{\underline{\omega}^2}{2s} & s + \frac{\underline{\omega}^2}{2s} \end{array} \right)
\end{equation}
and $\underline{\omega} = (\omega_y,\omega_z)$.  
The matrix $M(s,\underline{\omega})$ has the eigenvalues 
$\pm\sqrt{s^2 + \underline{\omega}^2}$.

We first look at the case $\underline{\omega} = 0$, which corresponds
to solutions that have trivial $y$ and $z$ dependence. For those, the
matrix $M(s,\underline{\omega})$ is diagonal and since $\mbox{Re}(s)
> 0$ we see that we must have $\tilde{u}_{ij}^{(+)} = 0$ for the
solution to be in $L^2$. The boundary conditions
(\ref{Eq:CPBC}), (\ref{Eq:Free}) yield
\begin{equation}
\tilde{u}_{xx}^{(-)}(0) = 0, \qquad
\hat{\tilde{u}}_{AB}^{(-)}(0) = 0,\qquad 
s\delta^{AB}\tilde{u}_{AB}^{(-)}(0) = 0,\qquad
s \tilde{u}_{xA}^{(-)}(0) = 0,
\end{equation}
therefore we have only the trivial solution.  There are no ill posed
modes with trivial dependence on the variables that are tangential to
the boundary. We show now that the situation becomes rather more
complicated when one considers modes that have a nontrivial tangential
dependence.

Assume that $\underline{\omega} \neq 0$. Following the analysis of
section \ref{eigenvalues} we introduce a unitary matrix $U =
U(s,\underline{\omega})$ that brings the matrix
$M(s,\underline{\omega})$ into upper triangular form.  To lighten the
notation we introduce the quantities $\zeta = s/|\underline{\omega}|$,
$\lambda = \sqrt{1 + \zeta^2}$, $\psi(\zeta) = (\lambda-\zeta)^2$ and
$N = 1 + |\psi(\zeta)|^2$. One can then verify that the matrix (a star
denotes complex conjugation)
\begin{equation}
U(\zeta) = N^{-1/2}\left( \begin{array}{cc} -\frac{|\zeta|}{\zeta} &
  \frac{|\zeta|}{\zeta^*}\psi^* \\
  \frac{|\zeta|}{\zeta}\psi & \frac{|\zeta|}{\zeta^*} \end{array} \right)
\end{equation}
is unitary and satisfies
\begin{equation}
U(\zeta)^{-1} M U(\zeta) = |\underline{\omega}| 
\left( \begin{array}{cc} -\lambda & M_0(\zeta) \\
 0 & \lambda \end{array} \right), \qquad
M_0(\zeta) = \frac{1 + \psi^*(2 + 4\zeta^2 + \psi^*)}{2\zeta^* N}\, ,
\end{equation}
where in $\lambda$ the branch is chosen such that for
$\mbox{Re}(\zeta) > 0$, $\mbox{Re}(\lambda) > 0$.  In terms of the new
variables $(v^{(-)}_{ij},v^{(+)}_{ij})^T =
U^{-1}(\tilde{u}^{(-)}_{ij},\tilde{u}^{(+)}_{ij})^T $
Eq. (\ref{Eq:ODE}) yields
\begin{eqnarray}
|\underline{\omega}|^{-1} \partial_x v_{ij}^{(-)} &=& -\lambda v_{ij}^{(-)}
+ M_0(\zeta) v_{ij}^{(+)}, \\
|\underline{\omega}|^{-1}\partial_x v_{ij}^{(+)} &=& \lambda v_{ij}^{(+)}.
\end{eqnarray}
For the solution to be in $L^2$, we must have $v_{ij}^{(+)} = 0$.
This implies that $v_{ij}^{(-)} = e^{-\lambda |\underline{\omega}| x}
\sigma_{ij}$, where 
$\sigma_{ij}$ are constants which describe the value that
$v_{ij}^{(-)}$ takes at the boundary. Using the matrix $U(\zeta)$ we can
express the $\tilde{u}$ variables at the boundary as
\begin{eqnarray}
\tilde{u}_{ij}^{(-)}(0) &=& -N^{-1/2}\frac{|\zeta|}{\zeta} \sigma_{ij}\, ,
\label{Eq:utildein}\\
\tilde{u}_{ij}^{(+)}(0) &=& N^{-1/2}\frac{|\zeta|}{\zeta}\psi(\zeta) \sigma_{ij}\, ,
\label{Eq:utildeout}\\
\tilde{u}_{Aij}^{(0)}(0) &=& \frac{i}{\sqrt{2}} 
 \frac{\hat{\omega}_A}{\zeta} N^{-1/2}\frac{|\zeta|}{\zeta}\left( 1-\psi(\zeta) \right)\sigma_{ij}\, ,
\label{Eq:utildezero}
\end{eqnarray}
where $\hat{\omega}_A = \omega_A / |\underline{\omega}|$.

Using Eqs. (\ref{Eq:SSEC1}), (\ref{Eq:utildein}) and (\ref{Eq:utildeout})
in the boundary condition (\ref{Eq:Free}), we find that
\begin{equation}
(1 + c\psi(\zeta)) \sigma_{xx} = 0, \qquad
(1 + d\psi(\zeta)) \hat{\sigma}_{AB} = 0,
\end{equation}
where $\hat{\sigma}_{AB}$ denotes the tracefree part of $\sigma_{AB}$.
Since the function $\psi(\zeta)$ maps the half plane $\mbox{Re}(\zeta) > 0$ to
the interior of the unit circle and since $|c|\leq 1$, $|d|\leq 1$, it
follows that $\sigma_{xx} = 0$ and $\hat{\sigma}_{AB} = 0$.

Next, we insert all of this into the boundary conditions
(\ref{Eq:CPBC}). The result is more conveniently expressed if one
introduces a normalized vector $\hat{\xi}_A$ that is orthogonal to
$\hat{\omega}_A$ and considers the components $\sigma_{x\omega} =
\delta^{AB}\sigma_{xA} \hat{\omega}_B$ and $\sigma_{x\xi} =
\delta^{AB}\sigma_{xA} \hat{\xi}_B$. The projection of the second
equation in (\ref{Eq:CPBC}) along $\hat{\xi}$ implies that
$\sigma_{x\xi}$ must vanish, while the remaining equations in
(\ref{Eq:CPBC}) imply that $\sigma \equiv \delta^{AB}\sigma_{AB}$ and
$\sigma_{x\omega}$ must satisfy the following $2\times 2$ system:
\begin{equation}
L_-(\zeta)\left( \begin{array}{c} \sigma \\ \sigma_{x\omega} \end{array} \right) = 0,\qquad
L_-(\zeta) = \left( \begin{array}{cc} 2\lambda(1 + a\psi) - \kappa(1+a)(\lambda-\zeta) & 2i(1+a\psi) + i\kappa(1+a)(1+\psi) \\ 
 i(1+b\psi) - i\kappa(1+b)(1+\psi) & 2\lambda(1+b\psi) + 2\kappa(1+b)(\lambda-\zeta) \end{array} \right).
\label{Eq:DetEC}
\end{equation}
The determinant of $L_-(\zeta)$ is given by
\begin{equation}
\det L_-(\zeta) = (6+4\zeta^2)\left[ (1+a\psi(\zeta))(1+b\psi(\zeta)) - \kappa^2(1+a)(1+b)\psi(\zeta) \right].
\end{equation}
Clearly, the first factor cannot be zero since $\mbox{Re}(\zeta) > 0$.
Therefore, $\det L_-(\zeta)$ can only vanish if the term inside the square brackets
does.

We now focus on the different cases discussed in the previous section:
\begin{enumerate}
\item $a=-1$, $b=1$, $c=d=1$:\\
In this case, the second term inside the square brackets vanishes and
the first term is never zero since $|\psi(\zeta)| < 1$. Therefore, the
resulting formulation possesses no ill posed modes.  Of course, when
$0 < \eta < 2$ this is consistent with our calculation in
Ref. \cite{CPSTR} where the estimates we have derived exclude the
presence of ill posed modes.
\item $a=1$, $b=-1$, $c=d=-1$:\\
The result is the same as in the previous case.
\item $a=0$, $b=0$:\\
In this case, the terms inside the square brackets simplify to $1 -
\kappa^2\psi(\zeta)$. A small calculation reveals that this can only
be zero if $\zeta = (\kappa^2-1)/2|\kappa|$ and $\kappa\neq
0$. Therefore, $\det L_-(\zeta)$ has a zero with $\mbox{Re}(\zeta) > 0$
if and only if $\kappa^2 > 1$.  This is equivalent to $\eta < 0$ or
$\eta > 8/3$.  Therefore, setting the ingoing constraint variables to
zero in the family of generalized Einstein-Christoffel systems does
indeed yield ill posed boundary conditions if the parameter $\eta$
lies outside the interval $[0,8/3]$.
\item $a=0$, $b=1$:\\ 
Here the terms inside the square brackets reduce to $1 + (1 -
2\kappa^2)\psi(\zeta)$. Since the function $\psi$ maps the positive real
axis onto the open interval $(0,1)$ it follows that this expression
never vanishes if and only if $\kappa^2\leq 1$.  In particular, one
has ill posed modes when $\eta=4$ and the boundary conditions that
were proposed in Ref. \cite{FG2} yield, at least when linearized
around flat spacetime, an ill posed initial-boundary formulation.  On
the other hand, if at the boundary one considers the equations $G_{xy}
= G_{xz} = 0$ and the equation $G_{xt} = 0$ instead of the combination
$G_{xx} - G_{xt} = 0$ one has $a=-1$ and the resulting formulation
does not in fact suffer from possessing ill posed modes.
\end{enumerate}

%%%%%%%%%%%%%%%%%%%%%%%%%%%%%%%%%%%%%%%%%%%%%%%%%%%%%%%%%%%%%%
\section{Violations of the constraints}
\label{CV}
%%%%%%%%%%%%%%%%%%%%%%%%%%%%%%%%%%%%%%%%%%%%%%%%%%%%%%%%%%%%%%

In this section, we show that the ill posed modes we have found in the
previous section violate the constraints. In order to see this, we use
these ill posed modes to compute the constraint variables $C_j$. From
$K_{ij} = (u_{ij}^{(-)} + u_{ij}^{(+)})/\sqrt{2}$,
Eqs. (\ref{Eq:Mom}), (\ref{Eq:utildein}), (\ref{Eq:utildeout}) and
$\sigma_{xx} = 0$, $\hat{\sigma}_{AB} = 0$, we have
\begin{eqnarray}
C_x &=& -\frac{|\underline{\omega}|}{\sqrt{2N}}\frac{|\zeta|}{\zeta} (1-\psi(\zeta))
 \left( \lambda\sigma + i\sigma_{x\omega} \right)
 \exp\left[ |\underline{\omega}|\left(\zeta t - \lambda x + i\hat{\omega}_A x^A \right) \right],
\label{Eq:Cxill}\\
\omega^A C_A &=& \frac{|\underline{\omega}|}{\sqrt{8N}}\frac{|\zeta|}{\zeta} (1-\psi(\zeta))
 \left( i\sigma + 2\lambda\sigma_{x\omega} \right) 
 \exp\left[ |\underline{\omega}|\left(\zeta t - \lambda x + i\hat{\omega}_A x^A \right) \right],
\label{Eq:CAill}
\end{eqnarray}
where $(\sigma,\sigma_{x\omega})$ is a nontrivial solution to Eq.
(\ref{Eq:DetEC}).
Since $|\psi(\zeta)| < 1$ for $\mbox{Re}(\zeta) > 0$, and since
\begin{equation}
\det\left( \begin{array}{cc} \lambda & i \\ i & 2\lambda \end{array}\right) 
 = 3 + 2\zeta^2 \neq 0,
\end{equation}
we see that the variables $C_x$ and $\omega^A C_A$ cannot
simultaneously vanish. Therefore, all the ill posed modes we have
found are {\it constraint violating} modes. This means that under
generic small perturbations of the initial data these modes will be
excited and the constraint variables will grow exponentially with an
exponential factor that can be arbitrarily large. In this sense, the
boundary conditions that lead to ill posed modes do {\it not} preserve
the constraints. We point out that the constraint variables
constructed from any solution of the main evolution system
(\ref{Eq:Kij}), (\ref{Eq:fkij}) with boundary conditions
(\ref{Eq:CPBC}), (\ref{Eq:Free}) provide a solution of the evolution of
the constraint variables,
Eqs. (\ref{Eq:C}), (\ref{Eq:Cj}), (\ref{Eq:Tij}), (\ref{Eq:Vij}) with
boundary conditions (\ref{Eq:CPBC}). Since we have shown that the
constraint variables constructed from ill posed modes are ill posed
modes themselves (see Eqs. (\ref{Eq:Cxill}), (\ref{Eq:CAill})), the
IBVP for the constraint variables cannot be well posed. This
emphasizes the importance of looking at the evolution system for the
constraints and checking its well posedness when deriving CPBC for
Einstein's equations.

We conclude this section with two remarks. First, one can check that
the evolution system for the constraint variables, Eqs.
(\ref{Eq:C}), (\ref{Eq:Cj}), (\ref{Eq:Tij}), (\ref{Eq:Vij}), is strongly
hyperbolic for any nonvanishing value of the parameter $\eta$. On the
other hand, our analysis above shows the existence of ill posed modes
when $\eta$ lies outside of the interval $[0,8/3]$ and the coupling
constants $a$ and $b$ are chosen as in the cases 3. and 4. of the
previous section. This illustrates that applying maximal dissipative
boundary conditions to evolution systems that are strongly hyperbolic
(but not symmetrizable) does not necessarily yield a well posed
problem.

The second remark concerns the choice $a=b=-1$ for the coupling
constants in Eq. (\ref{Eq:CPBC}). In this case it follows that the
determinant condition is always satisfied, regardless of the value for
the parameter $\eta$. In fact, one can show that the resulting
boundary conditions are constraint preserving: The evolution equations
imply that the constraint variable $C_j$ satisfies the wave equation:
\begin{equation}
\partial_t^2 C_j = \delta^{rs}\partial_r\partial_s C_j\; .
\end{equation}
On the other hand, the choice $a=b=-1$ corresponds to imposing the
momentum constraint at the boundary. Since $C_j$ satisfies the wave
equation, this implies that $C_j = 0$ everywhere, if $C_j$ is
satisfied initially. It then follows from Eqs. (\ref{Eq:C}),
(\ref{Eq:Tij}), (\ref{Eq:Vij}) that the remaining constraints are also
satisfied if they are satisfied initially. This explains why one has
CPBC for all $\eta \neq 0$ when $a=b=-1$. However, the above
argumentation is expected to  break down when one considers the nonlinear
regime since in this case lower order terms might prevent one from
obtaining a closed system for $C_j$ alone. In this case, one has to
rely on the symmetrizer for the system
(\ref{Eq:C}), (\ref{Eq:Cj}), (\ref{Eq:Tij}), (\ref{Eq:Vij}) which was
constructed in Ref. \cite{CPSTR}, and one might not be able to show
that the constraints propagate when $\eta$ lies outside the interval
$[0,2)$, even when $a=b=-1$.

%%%%%%%%%%%%%%%%%%%%%%%%%%%%%%%%%%%%%%%%%%%%%%%%%%%%%%%%%%%%%%
\section{Conclusions}
\label{conclusions}
%%%%%%%%%%%%%%%%%%%%%%%%%%%%%%%%%%%%%%%%%%%%%%%%%%%%%%%%%%%%%%

We have analyzed ill posed modes in the family of the generalized
Einstein-Christoffel formulation of Einstein's equations with
boundaries. We considered boundary conditions which result from
coupling the ingoing characteristic constraint variables to the
outgoing ones. Specifically, the cases we have studied include the
boundary conditions we have obtained in Ref. \cite{CPSTR} and the
boundary conditions that originate from considering the projection of
Einstein's equations along the normal to the boundary. When linear
fluctuations around Minkowski space are considered, we have shown that
the formulation is subject to constraint violating ill posed modes
unless the parameters in the equations and the coupling between the
in- and outgoing constraint variables are chosen carefully. In fact,
it is not difficult to show that if the coupling constants $a$ and $b$
are real and satisfy $-1 < a \leq 1$ and $-1 < b \leq 1$ there are
always ill posed modes as long as the parameter $\eta$ lies outside
the interval $[0,8/3]$. In particular, this is the case when the
ingoing constraint variables are set to zero. Furthermore, there are
ill posed modes for the boundary conditions that were obtained in
Ref. \cite{FG2} when applied to the linearized Einstein-Christoffel
system ($\eta = 4$). However, our analysis also reveals that these ill
posed modes could easily be avoided by imposing a different linear
combination of Einstein's equations at the boundary or by changing the
parameter $\eta$ such that it lies in the interval $0 < \eta \leq
8/3$. In any case, our analysis highlights the importance of studying
the evolution system for the constraint variables and ensuring its
well posedness since all the ill posed modes we have found are
constraint violating. In particular, the formulations we have studied
in this article show that even though the main evolution system is
symmetric hyperbolic, the evolution equations for the constraint
variables is not necessarily symmetrizable. For the cases in which the
propagation of the constraints is described by a system that is
strongly hyperbolic (but not symmetrizable) we have shown that
specifying maximal dissipative boundary conditions can lead to an ill
posed system.

It is interesting to note that all the ill posed modes that appear
have a nontrivial dependence in the spatial directions that are
tangential to the boundary surface. Therefore, such modes would not be
present in the one-dimensional case. This might explain why the
numerical simulations in Ref. \cite{CLT}, where the
Einstein-Christoffel system ($\eta=4$) was evolved using boundary
conditions obtained by setting the ingoing constraints to zero, did
not show any ill posed modes.

The simple analytic method we have used in this article, which is
based on the determinant condition (\ref{Eq:Det}), should be used to
test the well posedness of the boundary conditions before numerically
evolving any evolution system since the presence of ill posed modes
would detrimentally affect numerical stability. However, we also
stress that more work is required to derive sufficient conditions for
well posedness for the choices of parameters when the determinant
condition is satisfied.  In particular, it would be worthwhile to 
analyze CPBC where the incoming physical variables can be freely
specified.

%%%%%%%%%%%%%%%%%%%%%%%%%%%%%%%%%%%%%%%%%%%%%%%%%%%%%%%%%%%%%%
\section*{Acknowledgments}
%%%%%%%%%%%%%%%%%%%%%%%%%%%%%%%%%%%%%%%%%%%%%%%%%%%%%%%%%%%%%%
We particularly thank Gabriel Nagy for many valuable discussions. We
also thank L. Lehner, J. Pullin, O. Reula and M. Tiglio for useful
comments. This work was supported in part by grants NSF-PHY-9800973,
NSF-INT-0204937, and the Horace C. Hearne Jr. Institute of Theoretical
Physics.

%%%%%%%%%%%%%%%%%%%%%%%%%%%%%%%%%%%%%%%%%%%%%%%%


\begin{thebibliography}{10}
%%%%%%%%%%%%%%%%%%%%%%%%%%%%%%%%%%%%%%%%%%%%%%%%%

\bibitem{R}
O. Reula, Living Reviews in Relativity {\bf 1}, 3 (1998).

\bibitem{L}
L. Lehner, Class. Quantum Grav. {\bf 18}, R25 (2001).

\bibitem{FN} 
H. Friedrich and G. Nagy, Comm. Math. Phys. {\bf 201}, 619 (1999).

\bibitem{Stewart} 
J.M. Stewart, Class. Quantum Grav. {\bf 15}, 2865 (1998).

\bibitem{IR} 
M.S. Iriondo and O.A. Reula, Phys. Rev. D {\bf 65}, 044024 (2002).

\bibitem{CLT} 
G. Calabrese, L. Lehner, and M. Tiglio, 
Phys. Rev. D {\bf 65}, 104031 (2002).

\bibitem{CPSTR}
G. Calabrese, J. Pullin, O. Sarbach, M. Tiglio, and O. Reula,
{\em ``Well posed constraint preserving boundary conditions for the linearized Einstein equations,''}
arXiv:gr-qc/0209017.

\bibitem{SSW} 
B. Szilagyi, B. Schmidt, and J. Winicour,
Phys. Rev. D {\bf 65}, 064015 (2002).

\bibitem{SW} 
B. Szilagyi and J. Winicour,
{\em ``Well Posed Initial-Boundary Evolution in General Relativity,''}
arXiv:gr-qc/0205044.

\bibitem{FG1}
S. Frittelli and R. Gomez,
{\em ``Einstein boundary conditions for the 3+1 Einstein equations,''}
arXiv:gr-qc/0302071.

\bibitem{FG2}
S. Frittelli and R. Gomez,
{\em ``Boundary conditions for hyperbolic formulations of the Einstein equations,''}
arXiv:gr-qc/0302032.

\bibitem {BB}
J.M. Bardeen and L.T. Buchman,
Phys. Rev. D {\bf 65}, 064037 (2002).

\bibitem{GKO-Book}
B. Gustafsson, H. Kreiss, and J. Oliger,
{\em ``Time dependent problems and difference methods,''}
John Wiley \& Sons, New York (1995).

\bibitem{CPST-Convergence}
G. Calabrese, J. Pullin, O. Sarbach and M. Tiglio,
Phys.\ Rev.\ D {\bf 66}, 041501 (2002).

\bibitem{LaxPh} 
P.D. Lax, and R.S. Phillips, 
Commun. Pure Appl. Math. {\bf 13}, 427 (1960).

\bibitem{Secchi}
P. Secchi, Diff. Int. Eq. {\bf 9}, 671 (1996); 
Arch. Rat. Mech. Anal. {\bf 134}, 595 (1996).

\bibitem{Rauch} 
J. Rauch, Trans. Am. Math. Soc. {\bf 291}, 167 (1985).

\bibitem{Kreiss}
H. Kreiss,
Commun. Pure Appl. Math. {\bf 23}, 277 (1970).

\bibitem{MO}
A. Majda and S. Osher,
Commun. Pure Appl. Math. {\bf 28}, 607 (1975).

\bibitem{KST}
L.E. Kidder, M.A. Scheel, and S.A. Teukolsky,
Phys. Rev. D {\bf 64}, 064017 (2001).

\bibitem{AY}
A. Anderson and J.W. York, Jr., 
Phys. Rev. Lett. {\bf 82}, 4384 (1999).

\bibitem{FR}
S. Frittelli and O.A. Reula,
Phys. Rev. Lett. {\bf 76}, 4667 (1996). 

\bibitem{KL-Book} 
H.O. Kreiss, J. Lorenz,
{\em ``Initial-Boundary Value Problems and the Navier-Stokes Equations,''} 
Academic Press, (1989).


\end{thebibliography}
\end{document}